\documentclass[aps,prd,showpacs,onecolumn,superscriptaddress,nofootinbib,floatfix]{revtex4}
\usepackage{amsmath}
\usepackage{amssymb,amsfonts}
\usepackage{graphicx}

\def\be{\begin{equation}}
\def\ee{\end{equation}}
\def\ba{\begin{eqnarray}}
\def\ea{\end{eqnarray}}

\def\g{\gamma}

\newcommand{\mubar}{{\bar \mu}} % mu bar

\newcommand{\Ham}{{\mathcal H}}

\begin{document}

\title{Magnetic Bianchi I Universe in Loop Quantum Cosmology}

\author{Roy Maartens}
\email{roy.maartens@port.ac.uk} \affiliation{Institute of
Cosmology \& Gravitation,
             University of Portsmouth, Portsmouth PO1 2EG, UK}
\author{Kevin Vandersloot}
\email{kevin.vandersloot@port.ac.uk} \affiliation{Institute of
Cosmology \& Gravitation,
             University of Portsmouth, Portsmouth PO1 2EG, UK}

\date{\today}

\begin{abstract}

We examine the dynamical consequences of homogeneous cosmological
magnetic fields in the framework of loop quantum cosmology. We
show that a big-bounce occurs in a collapsing magnetized Bianchi I
universe, thus extending the known cases of singularity-avoidance.
Previous work has shown that perfect fluid Bianchi I universes in
loop quantum cosmology avoid the singularity via a bounce. The
fluid has zero anisotropic stress, and the shear anisotropy in
this case is conserved through the bounce. By contrast, the
magnetic field has nonzero anisotropic stress, and shear
anisotropy is not conserved through the bounce. After the bounce,
the universe enters a classical phase. The addition of a dust
fluid does not change these results qualitatively.

\end{abstract}

%\pacs{04.60.Pp, 04.70.Bw, 98.80.Qc, 03.65.Sq}

\maketitle

\section{Introduction}

Loop quantum cosmology (LQC) is a theory of quantum cosmology
based on the more general theory of loop quantum
gravity~\cite{Bojowald:2006da} (for reviews, see~\cite{LQG,
Rovelli, Thiemann}). One of the most important predictions of LQC
is the avoidance of the big-bang singularity, which is replaced by
a bouncing universe~\cite{BBLQC1, BBLQC2, BBLQC3, BBLQC4, BBLQC5}
for isotropic models sourced by a massless scalar field. This
result has been derived rigorously at the level of the quantum
theory, but has also been understood at the level of approximate
effective classical equations that capture the main features of
the quantum dynamics.

Extending the rigorous quantum dynamics to the anisotropic Bianchi
I model is challenging. Because of this, the dynamics of Bianchi I
in LQC have mostly been studied by extrapolating the approximate
effective semi-classical equations that proved successful in the
isotropic case. For matter with zero anisotropic stress, the
effective equations predict a bounce that avoids the classical
singularity~\cite{Chiou:2007sp}, thus extending the results of the
isotropic case. Furthermore, it was shown that the shear
anisotropy does not blow up in the collapsing phase, but remain
finite through the bounce. Several ambiguities in the quantum
construction of the Hamiltonian constraint were
considered~\cite{Chiou:2007sp}, one of which has since been
favored by gauge considerations~\cite{Chiou:2007mg} and by a more
thorough construction of the quantum theory~\cite{AWE}. With this
choice of quantization scheme, it was shown that the anisotropic
shear is in fact conserved across the bounce when the matter has
zero anisotropic stress~\cite{Chiou:2007sp}.

In this paper, we couple a homogeneous magnetic field to a Bianchi
I universe and consider the effective semi-classical modifications
to the equations of motion. We show that the singularity is still
avoided via a bounce, during which anisotropies remain finite.
However, the anisotropic stress in the magnetic field leads to a
non-conservation of shear anisotropy through the bounce, in
contrast to the case where of matter has zero anisotropic stress.

\section{Classical Equations}
\label{cleq}

The inclusion of a cosmological magnetic field breaks isotropy, so
we consider an anisotropic Bianchi cosmology, the simplest being
the Bianchi I model:
 \ba
ds^2 = - dt^2 + a_1^2(t) \,dx^2 + a_2^2(t)\, dy^2 + a_3^2(t)
\,dz^2\,,~~ H_i:={\dot{a}_i \over a_i}\,.
 \ea
Maxwell's equations are
 \ba \label{Meq}
\nabla_{[\mu}F_{\nu \alpha]} = 0\,, \qquad \nabla_{\nu} F^{\mu
\nu} = J^{\mu}\,,
 \ea
where $J^\mu$ is the four-current. The Faraday tensor defines
electric and magnetic fields relative to observers with
four-velocity $u^\mu$~\cite{Barrow:2006ch}
 \be
E_\mu=F_{\mu\nu}u^\nu\,,~~~ B_\mu= \frac{1}{2} \varepsilon_{\mu
\nu \alpha} F^{\nu \alpha}\,,
 \ee
where $ \varepsilon_{\mu \nu \alpha}$ is the alternating tensor in
the observer's rest-space.

We assume high conductivity in the early universe, so that the
electric field is effectively zero, and Maxwell's equations reduce
to
 \ba
h_\mu{}^\nu\dot{B}_\nu &=& \left( \sigma_{\mu\nu}-{2\over 3}\Theta
h_{\mu\nu} \right)B^\nu\,, \label{ind}\\ h^{\mu\nu}\nabla_\mu
B_\nu &= & 0\,,~~ \varepsilon_{\mu \nu \alpha}\nabla^\nu B^\alpha=
h_{\mu\nu}J^\nu \,,
 \ea
where $h_{\mu\nu}=g_{\mu\nu}+u_\mu u_\nu$ projects into the
rest-space, $\Theta$ is the volume expansion and $\sigma_{\mu\nu}$
is the shear:
 \ba
\Theta=H_1+H_2+H_3=3{\dot a \over a}\,, ~~ a^3:=a_1a_2a_3\,, ~~~~
\sigma_i{}^j=\delta_i{}^j \left({\dot{a}_j \over a_j}- {\dot a
\over a} \right).
 \ea

For a homogeneous magnetic field in Bianchi I, the divergence
constraint is automatically satisfied, and the curl constraint
shows that there is no 3-current. To solve the induction
equation~\eqref{ind}, we assume without loss of generality that
the magnetic field is aligned along the $x$-direction:
$B_\mu=B_1(t)\delta_\mu{}^1$. The solution is
 \ba \label{Bt}
B_\mu B^\mu={\beta^2 \over (a_2a_3)^2}\,,~~~ {B}^1 = {\beta \over
a^3}\,,
 \ea
where $\beta$ is constant, in agreement with~\cite{jac}.

The electromagnetic energy-momentum tensor is given by
 \ba
T^F_{\mu \nu} &=& - F_{\mu\alpha} F^\alpha{}_{\nu} - \frac{1}{4}
g_{\mu \nu} F_{\alpha\gamma} F^{\alpha\gamma}\nonumber\\
&=& \rho_B u_\mu u_\nu + {1\over3}\rho_B h_{\mu\nu}
+\pi^B_{\mu\nu}\,,
 \ea
where the magnetic energy density and anisotropic stress are
 \be \label{bra}
\rho_B={1\over2}B_\mu B^\mu \,,~~~ \pi^B_{\mu\nu}=
{1\over3}B_\alpha B^\alpha h_{\mu\nu}-B_\mu B_\nu\,.
 \ee

For the case with only a magnetic field, the Einstein field
equations $ G_{\mu\nu}=8\pi G T^F_{\mu\nu}$ lead to~\cite{jac}:
 \ba
H_1 + H_I &=& \frac{\gamma_{I}}{a^3}\,, ~~~ I=2,3\,, \label{eq1} \\
H_1^2 a^6 &=& \gamma_{2} \gamma_{3} - 4\pi G \beta^2a_1^2 \,,
 \ea
where $\gamma_{I}$ are constants. It follows that
 \ba \label{a1c}
\gamma_2\gamma_3>0\,, ~~~     a_1 \le a_{1m}= \frac{ \gamma_{2}
\gamma_{3}}{4\pi G \beta^2}\,,
 \ea
and thus any expansion in the magnetic field direction will
eventually come to rest at the maximum scale factor $a_{1m}$ and
turn around into a contracting phase. The solutions $a_i(t)$ can
be given analytically~\cite{jac}:
 \ba
a &\propto & (1 + f^2) f^{\pm \alpha - 1}\,,~~ f(a_1) :=
\frac{a_{1m}}{a_1} \left[ 1-\sqrt{1-
\left(\frac{a_1}{a_{1m}}\right)^2}\right],~~\alpha:=
\frac{\gamma_{2} + \gamma_{3}}{\sqrt{\gamma_{2}  \gamma_{3}}}\,,
\label{Vofa1}\\
a_I &\propto& (1 + f^2) f^{\pm \gamma_{I}/\sqrt{\gamma_{2}
\gamma_{3}}}\,, \label{a2ofa1}
 \ea
where the $\pm$ refers to the expanding (collapsing) branches
before (after) $a=a_{1m}$ is reached.

We can now analyze the singularity behavior of the general
solutions. From Eq.~\eqref{Vofa1}, the volume goes to zero or
infinity when $a_1$ goes to zero, depending on the value of
$\alpha$. There are two separate cases. The first is the
axisymmetric case, $a_2=a_3$, $\gamma_{2} = \gamma_{3},
\alpha=-2$. From Eqs.~\eqref{Vofa1} and \eqref{a2ofa1}, the
axisymmetric singularity is characterized by
 \ba \label{axising}
a_2= a_3  \rightarrow  \text{const}\,,~~~ a_1, a  \rightarrow  0
\,.
 \ea
Directions orthogonal to the magnetic field freeze as the
singularity is approached, while the magnetic field direction
contracts to zero, along with the total volume. This singularity
is present in all solutions for the axisymmetric case. The overall
evolution is characterized by expansion in the direction of the
magnetic field until $a_{1m}$ is reached. After that, the $a_1$
direction contracts and reaches the singularity in finite proper
time. For this type of evolution, past infinity is characterized
by $a, a_2, a_3 \rightarrow \infty$ while $a_1 \rightarrow 0$. In
addition, the time reversed scenario is possible.

The non-axisymmetric case is slightly more complicated, but again
all trajectories are singular. In this case $\gamma_{2} \neq
\gamma_{3}$, and the singularity is characterized by
 \ba
\label{nonaxising} a_2  \rightarrow 0\,, ~~~ a_3 \rightarrow
\infty \quad \text{if} \quad |\gamma_{2}| > |\gamma_{3}|\,, ~~~
a_1, a  \rightarrow  0\,,
 \ea
and vice-versa if $|\gamma_{2}| < |\gamma_{3}|$. Thus the
singularities are again given by $a_1$ going to zero, with one of
the orthogonal directions contracting to zero, while the other
expands to infinity in such a way that $a$ still goes to zero.

\section{Effective Loop Quantum Equations}

The loop quantum formulation is based on a Hamiltonian framework
where the gravitational degrees of freedom in the Bianchi I model
are encoded in three triad components $p_i$ and momentum
components $c_i$, related to the metric components as
 \be\label{eqn:p and a}
p_1=a_2a_3\,,~ p_2=a_1a_3\,,~ p_3=a_1 a_2\,,~~~c_i = \gamma
\dot{a}_i \,,
 \ee
where $\gamma$ is the real-valued Barbero-Immirzi parameter and
represents an ambiguity parameter of loop quantum gravity. Black
hole entropy calculations can be used to fix its value. In terms
of these variables, the Hamiltonian is given by
 \ba\label{Hamiltonian}
\Ham = \frac{- 1}{8 \pi G \g^2\sqrt{{p_1p_2p_3}}}
\left(c_2p_2c_3p_3\!+\!c_1p_1c_3p_3\!+\!c_1p_1c_2p_2\right)+
\,\Ham_{M},
 \ea
where $\Ham_{M}$ is the matter contribution to the Hamiltonian.
Einstein's equations can then be derived from Hamilton's
equations, which explicitly for this system are
 \be \label{doteqs}
\dot{p}_i = -8\pi G \g \frac{\partial \Ham }{ \partial c_i}\, ,
\qquad \dot{c}_i = 8\pi G \g \frac{\partial \Ham }{\partial
p_i}\,.
 \ee
The Hamiltonian must also vanish for the system:
 \be
\Ham=0 \,.
 \ee

The Hamiltonian for the matter contribution is proportional to the
energy density of the matter, so the magnetic Hamiltonian is given
by
 \be
\Ham_{B} = a^3\rho_B=\frac{a_1\beta^2}{2a_2a_3}\,.
 \ee
We will also consider a perfect fluid with constant equation of
state $w$. This can be added to the matter Hamiltonian by first
solving the conservation equation to give
 \be
\rho = Ca^{-3(1+w)}\,,~~~  \Ham_{\rm fluid} = C a^{-3w}\,,
 \ee
where $C$ is a constant.

Analyzing the system at the level of the quantum difference
equations of LQC for this model would be highly challenging, given
the complexity of the Bianchi I equations. We thus consider
approximate semi-classical equations of motion that incorporate
loop quantum modifications. These effective equations have been
shown to be very good approximations for the case of isotropic
cosmologies sourced by a massless scalar field~\cite{BBLQC1,
BBLQC2, BBLQC3, BBLQC4, BBLQC5}, and the results have been
extrapolated to more complicated models. The corrections modify
the general relativistic Hamiltonian~\eqref{Hamiltonian} to be of
the form~\cite{Chiou:2007sp}
 \ba \label{Heff}
\Ham_{\rm eff}= -\frac{1 }{8 \pi G \g^2 \sqrt{{p_1p_2p_3}}}
\left\{ \frac{\sin(\mubar_2 c_2) \sin(\mubar_3 c_3)}{\mubar_2
\mubar_3} \,p_2p_3 +\text{cyclic terms} \right\} + \Ham_{M}.
 \ea
The parameters $\mubar_i$ are the key ingredients determining the
quantum corrections. It is easy to see that in the limit $\mubar_i
\rightarrow 0$, the classical Hamiltonian~\eqref{Hamiltonian} is
recovered. The $\mubar_i$ parameters are assumed to be functions
of the triad components $p_i$, and their precise specification is
an ambiguity of the quantization. Two possible constructions are
discussed in~\cite{Chiou:2007sp}, although one of them has been
argued to be more physical on the grounds of certain gauge
invariance considerations in~\cite{Chiou:2007mg} and of a more
rigorous construction of the quantum theory in~\cite{AWE}. We will
focus on that scheme in this paper. The particular form is
 \be
\mubar_i = \frac{\sqrt{\Delta}}{a_i}\,,
 \ee
where $\Delta$ is a constant that is typically related to the
minimum area gap of loop quantum gravity. In this paper we assume
that $\Delta=O(1)$ in Planck units; the precise value will not
affect the qualitative results. The effective equations of motion
can be derived as in the general relativistic case, using
Hamilton's equations~\eqref{doteqs} and the vanishing of the
Hamiltonian.

\begin{figure}[ht]
\begin{center}
\includegraphics[width=7cm, keepaspectratio]
        {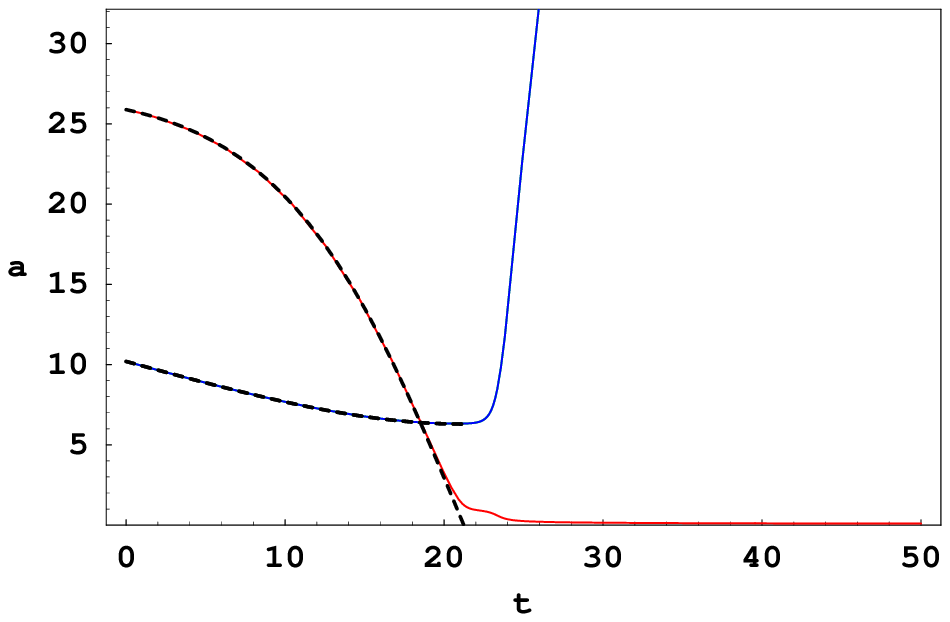}
\includegraphics[width=7cm, keepaspectratio]
        {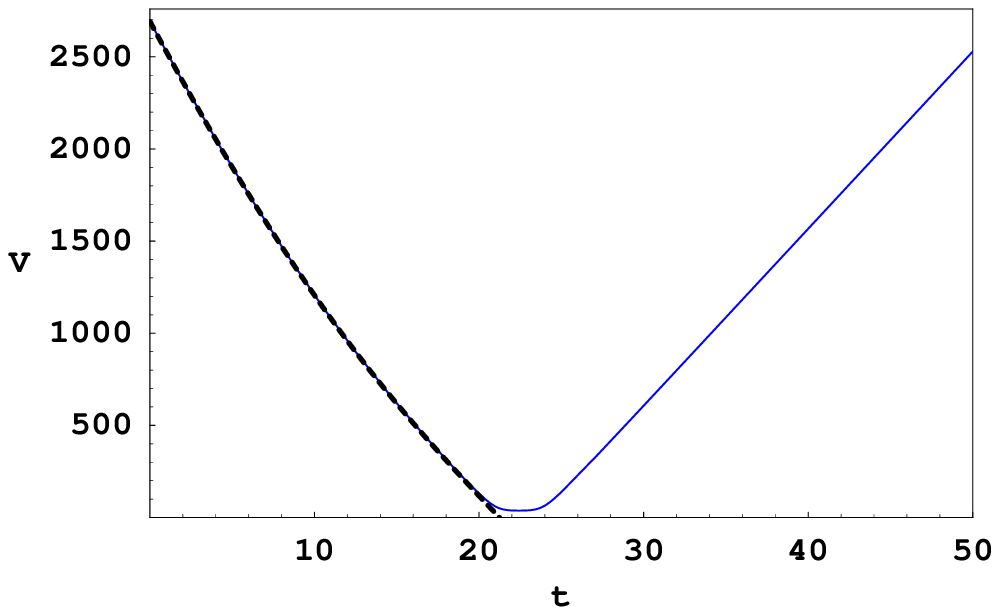}
\includegraphics[width=7cm, keepaspectratio]
        {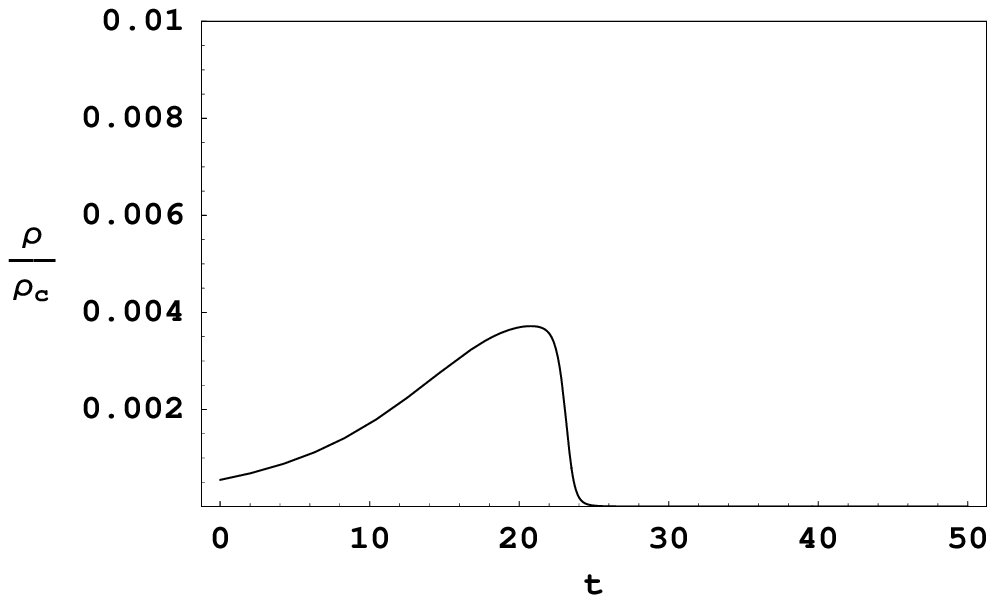}
\includegraphics[width=7cm, keepaspectratio]
        {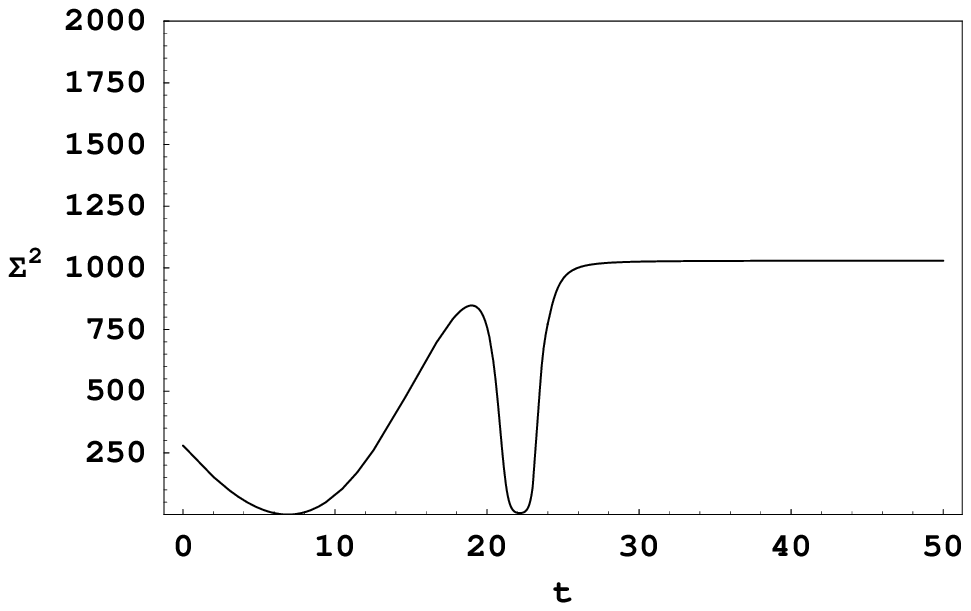}
\end{center}
\caption{The pure-magnetic axisymmetric case. {\em Top:} The scale
factors are shown on the left, and the volume factor $V=a_1a_2a_3$
is shown on the right. Dashed lines indicate the classical
behavior, $a_1\to0$, $a_2=a_3\to\,$const, and $V\to0$. Solid lines
show the effective loop quantum solutions. Quantum effects
regulate the singularity leading to a bounce in $a_2, a_3$ and
overall expansion of the universe. Note that $a_1$ continues to
decrease after the bounce. {\em Bottom:} The energy density as a
fraction of critical (left) and the shear energy density
$\Sigma^2=\sigma_{\mu\nu}\sigma^{\mu\nu}/2$ (right). This shows
the non-conservation of shear anisotropy through the bounce.}
\label{plot1}
\end{figure}

The equations of motion are sufficiently complicated to not allow
for an analytic solution. Despite that, some general conclusions
can be made from the form of the equations. First, as shown
in~\cite{Chiou:2007sp}, the vanishing of the
Hamiltonian~\eqref{Heff} immediately implies a bound on the energy
density of the matter. This arises from the bound in the $\sin$
terms of the constraint. The precise bound is the same critical
density that characterizes the bounce in the isotropic models:
 \be
\rho_c = \frac{3}{8 \pi G \g^2 \Delta} \,.
 \ee
The total energy density of the matter (magnetic plus fluid) must
be below this value. This is an indication that the classical
singularity (where the energy density diverges) is removed and
replaced by a bounce. The second conclusion from the effective
equations, is that if the matter has zero anisotropic stress, the
shear term is conserved before and after the bounce. Since the
magnetic field has non-zero anisotropic stress, Eq.~\eqref{bra},
this behavior is not guaranteed.

In the next section, we present numerical results.

\section{Numerical solution of the effective loop quantum equations}

\begin{figure}[htbp]
\begin{center}
\includegraphics[width=7cm, keepaspectratio]
        {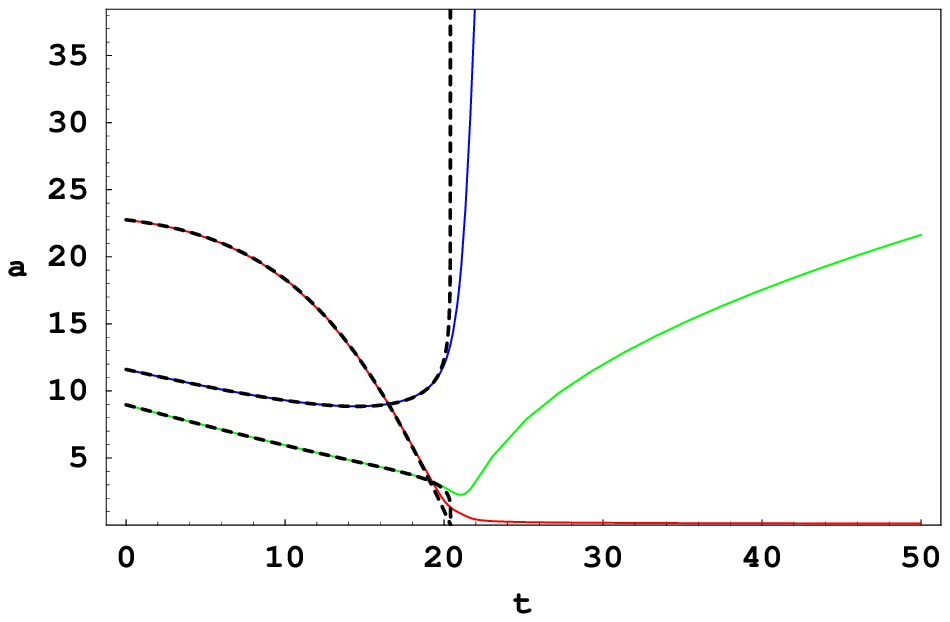}
\includegraphics[width=7cm, keepaspectratio]
        {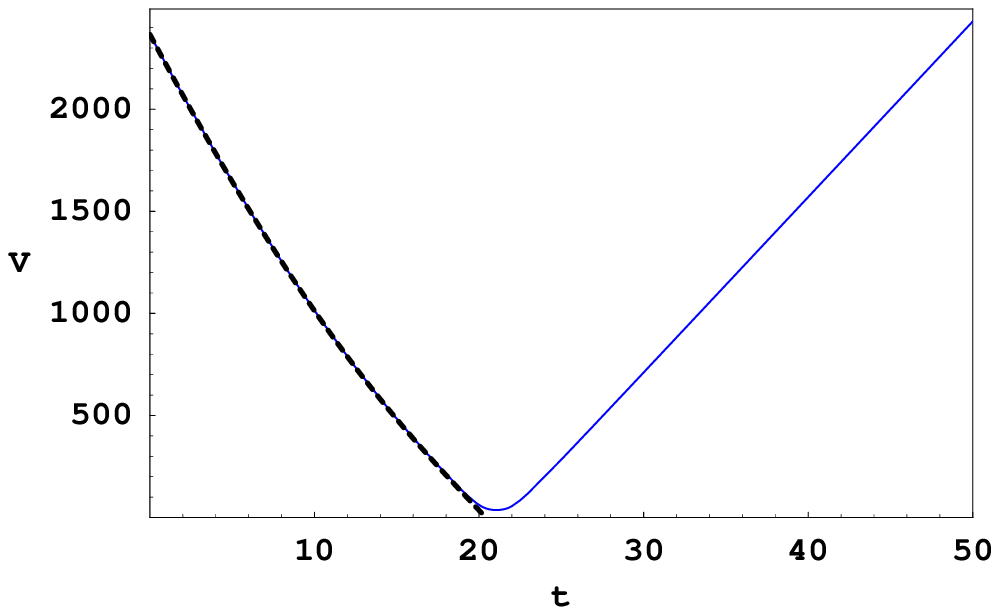}

\end{center}
\caption{The pure-magnetic non-axisymmetric case. Dashed lines
show the classical behavior, $a_1,a_2\to0$, $a_3\to\infty$ and
$V\to0$, and solid lines show the effective loop quantum
solutions. } \label{plot2}
\end{figure}

The first case is an axisymmetric spacetime ($a_2=a_3$) sourced
only by the magnetic field. Classically, the singularity is
characterized by Eq.~\eqref{axising}. We use initial conditions
corresponding to a classically collapsing universe approaching the
classical singularity. The solution is shown in Fig.~\ref{plot1}.
The quantum solution matches the classical well until the
singularity is approached. Then the quantum effects act
repulsively -- preventing $a_1$ from reaching zero, and leading to
bounces in $a_2= a_3$. The volume factor confirms that a bounce
replaces the classical singularity. The post-bounce expanding
universe has $a_2= a_3 \rightarrow \infty$ and $a_1 \rightarrow
0$, as in the classical case. Thus the quantum effects join a
classical contracting branch with an expanding classical branch.
The energy density shown in Fig.~\ref{plot1} remains bounded below
the classical critical density $\rho_c$ as expected from
analytical considerations. Finally, the shear energy density
$\Sigma^2=\sigma_{\mu\nu}\sigma^{\mu\nu}/2$ is shown to remain
finite through the evolution, but is not conserved through the
bounce. This is in contrast to the pure-fluid
case~\cite{Chiou:2007sp}, where shear is conserved. The difference
arises from the non-zero magnetic anisotropic stress,
Eq.~\eqref{bra}, which leads to production of shear anisotropy.

\begin{figure}[htbp]
\begin{center}
\includegraphics[width=7cm, keepaspectratio]
        {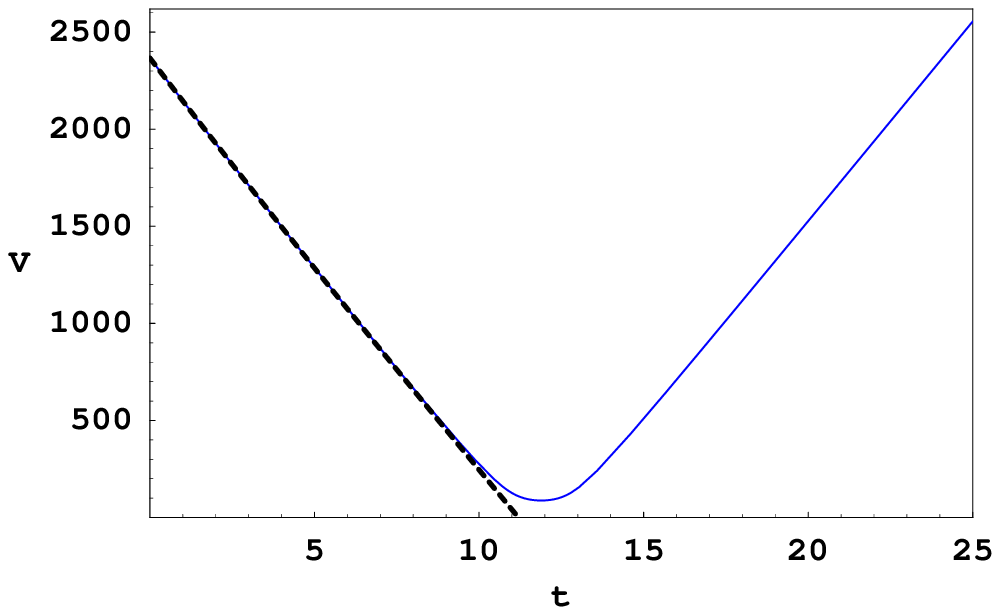}
\includegraphics[width=7cm, keepaspectratio]
        {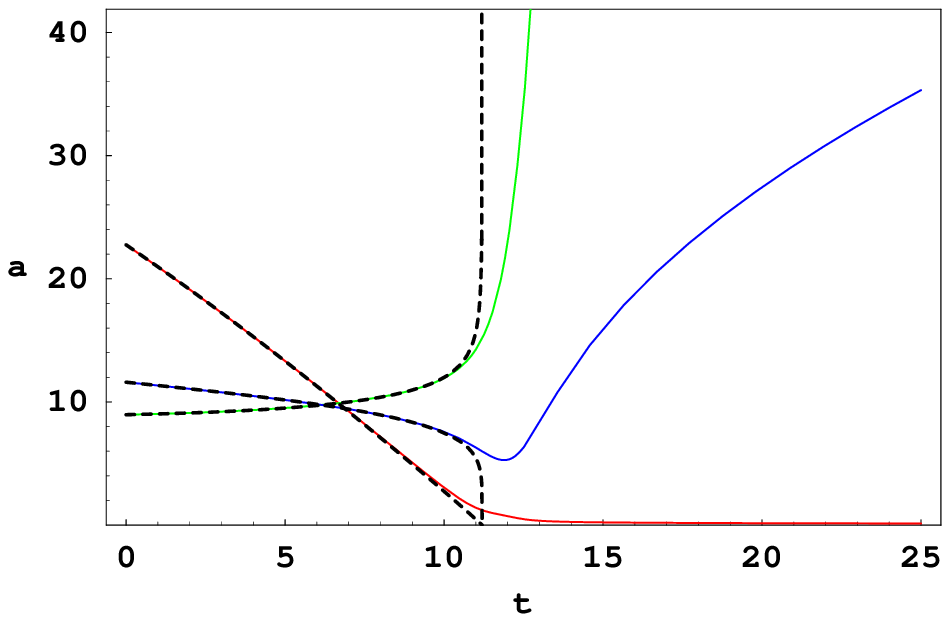}
\end{center}
\caption{As in Fig.~\ref{plot2}, but with different, Kasner-like
initial conditions.} \label{plot3}
\end{figure}

\begin{figure}[htbp]
\begin{center}
\includegraphics[width=7cm, keepaspectratio]
        {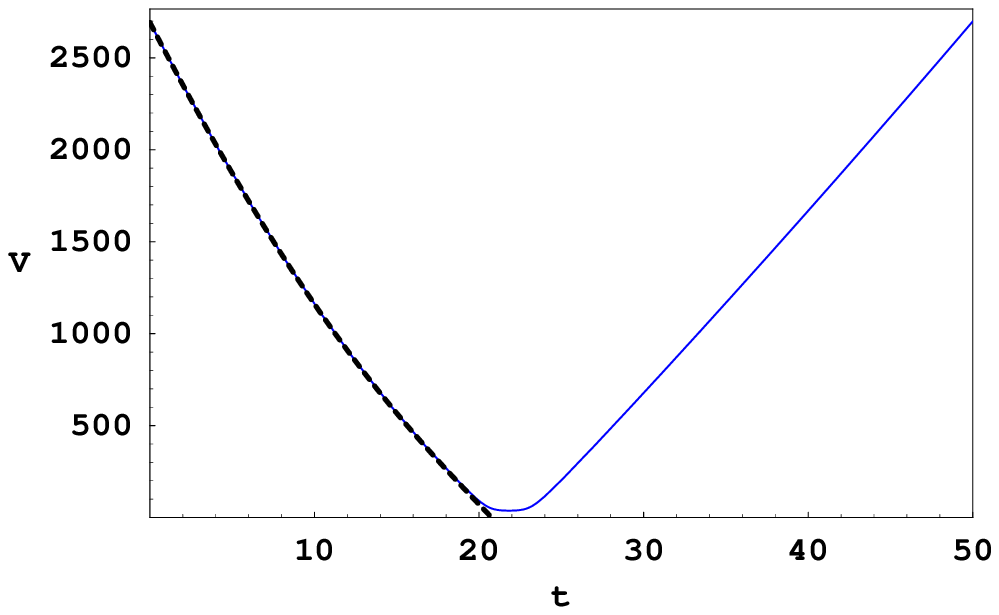}
\includegraphics[width=7cm, keepaspectratio]
        {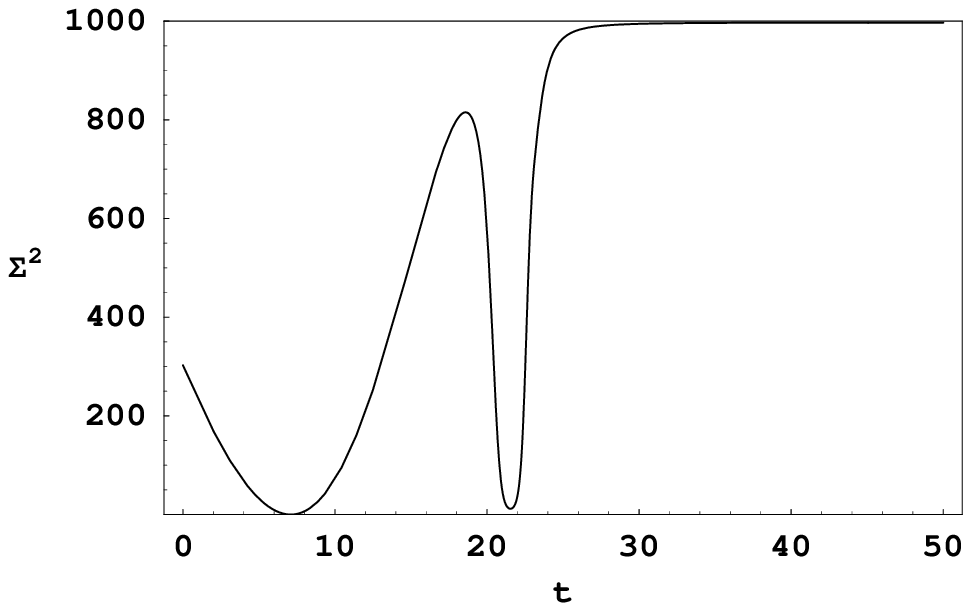}\hfill
\includegraphics[width=7cm, keepaspectratio]
        {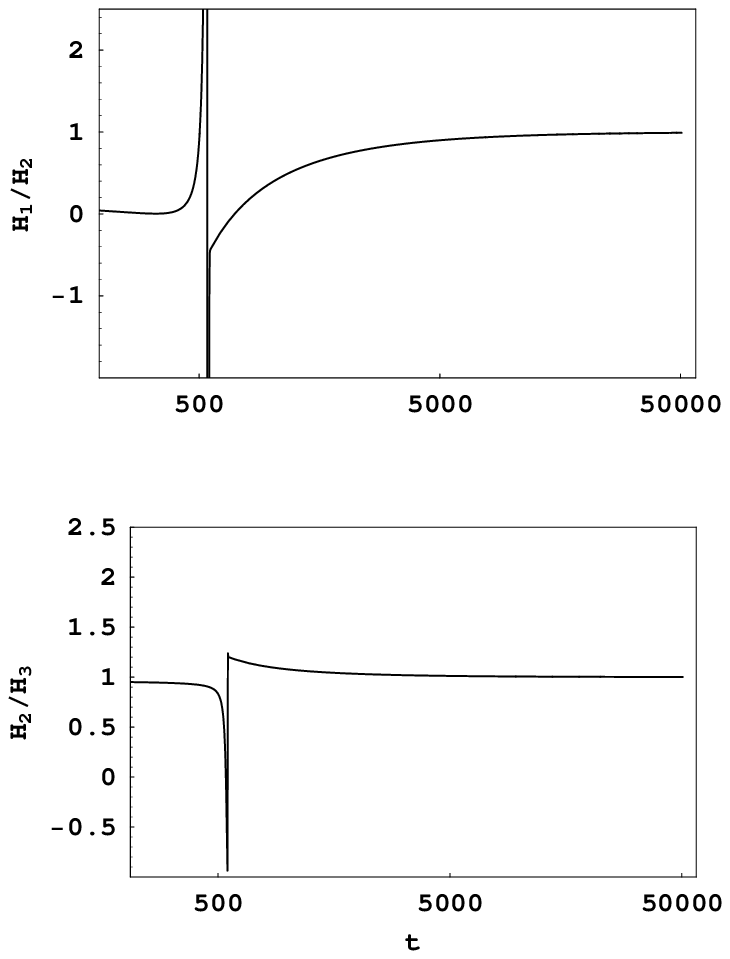}
\end{center}
\caption{Magnetic field and dust, non-axisymmetric. {\em Top:} The
classical singularity is avoided via a quantum bounce (left), and
the shear is not conserved through the bounce (right). {\em Middle
and bottom:} The ratios of expansion rates, showing the late-time,
post-bounce isotropization due to the dust. } \label{plot4}
\end{figure}

Figure~\ref{plot2} shows the solution for the pure-magnetic
non-axisymmetric case. The classical singularity is described in
Eq.~\eqref{nonaxising}. Once again a bounce occurs in the volume
near the point of the classical singularity. With the choice of
initial conditions in Fig.~\ref{plot2}, classically $a_3\to\infty$
while $a_2\to0$ at the singularity. With the quantum effects,
$a_1$ and $a_2$ are repelled from zero, and $a_2$ bounces. As in
the axisymmetric case, the post-bounce regime is an expanding
universe with $a_2, a_3$ expanding and $a_1$ contracting. The
overall behavior is qualitatively similar to the axisymmetric
case, and again the energy density of the magnetic field is
bounded below $\rho_c$ and the shear is not conserved.

An alternative non-axisymmetric choice of initial conditions is
shown in Fig.~\ref{plot3}. The initial conditions are chosen to be
in a Kasner phase, where two directions are contracting and one is
expanding, while overall there is contraction in the volume. The
qualitative behavior is qualitatively similar to the first
non-axisymmetric example in Fig.~\ref{plot2}.

As a final case, we included a dust perfect fluid, $w=0$, with the
magnetic field. We chose non-axisymmetric initial conditions with
the volume collapsing. Figure~\ref{plot4} shows a bounce
qualitatively similar to the pure-magnetic case. The shear term
remains finite, but is not conserved. At late times in the
post-bounce expansion phase, the dust begins to dominate the
evolution and the universe isotropizes, since the ratio of the
separate Hubble rates tends to one at late times.

\section{Conclusions}

We have extended the effective LQC treatment of Bianchi I
cosmologies by including a homogeneous magnetic field. We have
studied dynamics using the approximate effective equations of
motion that capture features of the (as yet unknown) true quantum
LQC dynamics. Thus our results are approximate, and a more
rigorous quantum construction would be needed to fully validate
them or provide additional corrections.

The effective equations indicate that the singularity-avoiding
bounce is not spoiled by the inclusion of a homogeneous magnetic
field. Extending the results of~\cite{Chiou:2007sp}, we showed
that shear anisotropy does not blow up as the classical
singularity is approached, but remains finite through the entire
evolution. In contrast to the pure-fluid case~\cite{Chiou:2007sp},
we showed shear is no longer conserved through the bounce, due to
the anisotropic stress carried by the magnetic field. Our results
indicate an interesting evolution of shear, possibly with a net
generation of shear, but further study is needed to check whether
this is generic. When we add a dust fluid to the magnetic field,
the qualitative behavior through the bounce is unchanged. However,
at late times after the bounce, the dominance of the dust ensures
that the universe isotropizes, unlike the pure-magnetic case.

\[ \]{\bf Acknowledgements:} KV was supported by the Marie Curie Incoming
International Grant M1F1-CT-2006-022239. The work of RM was
supported by the UK's STFC.

{}

\end{document}